\documentstyle[12pt,aaspp,tighten]{article}
\input BoxedEPS.tex
\SetRokickiEPSFSpecial
\HideDisplacementBoxes

\def\agt{>\kern-1.1em \lower1.1ex\hbox{$\sim$}\kern.3em}
\def\alt{<\kern-1.1em \lower1.1ex\hbox{$\sim$}\kern.3em}

\begin{document}

\title{A Balloon--Borne Millimeter--Wave Telescope for Cosmic Microwave
Background Anisotropy Measurements}
\author{ D.~J.~Fixsen\altaffilmark{1},
E.~S.~Cheng\altaffilmark{2},
D.~A.~Cottingham\altaffilmark{3},
W.~C.~Folz\altaffilmark{2},
C.~A.~Inman\altaffilmark{4},
M.~S.~Kowitt\altaffilmark{2},
S.~S.~Meyer\altaffilmark{4},
L.~A.~Page\altaffilmark{5},
J.~L.~Puchalla\altaffilmark{4},
J.~E.~Ruhl\altaffilmark{4},
R.~F.~Silverberg\altaffilmark{2}
}
\altaffiltext{1}{Applied Research Corporation, Code 685.3, NASA/GSFC,
Greenbelt MD 20771}
\altaffiltext{2}{NASA Goddard Space Flight Center, Code 685,
Greenbelt MD 20771}
\altaffiltext{3}{Global Science and Technology, Inc, Code 685.3, NASA/GSFC,
Greenbelt MD 20771}
\altaffiltext{4}{University of Chicago, Physics Dept., 5641 S. Ingleside,
Chicago IL 60637}
\altaffiltext{5}{Princeton University, Dept. of Physics, Jadwin Hall,
Princeton NJ 08540}

\begin{center}
\medskip
{Submitted to Astrophysical Journal, November 13, 1995}
\medskip
\end{center}

\begin{abstract}
We report on the characteristics and design details of
the Medium Scale Anisotropy Measurement (MSAM), a
millimeter-wave, balloon-borne telescope that has been used to observe
anisotropy in the Cosmic Microwave Background Radiation
(CMBR) on 0\fdg5 angular scales.
The gondola is capable of determining and maintaining absolute
orientation to a few arcminutes during a one-night flight.
Emphasis is placed on the optical and pointing
performance as well as the weight and power budgets.
We also discuss the total balloon/gondola mechanical system.  The
pendulation from this system is a ubiquitous perturbation on the
pointing system.
A detailed understanding in these areas is
needed for developing the next generation of balloon-borne instruments.
\end{abstract}


\section{Introduction}
The Medium Scale Anisotropy Measurement (MSAM)
is designed to measure variations in the 2.7~K Cosmic Microwave Background
Radiation (CMBR, \cite{mather94}) to 30 $\mu$K on 0\fdg5 angular scales.
The telescope we are describing was used for three
flights (from 1992 through 1995) during the first phase of the
project (MSAM1).
For these measurements, we used a bolometric receiver with four channels
at 5.6, 9.0, 16.5 and 22.5 cm$^{-1}$ (170, 270, 500, and 680 GHz or 1.8, 1.1,
0.61 and 0.44 mm) with approximately 1 cm$^{-1}$ bandwidth in each channel.
The results of the first flight (MSAM1-92, 1992 June 4--5) are reported by
\cite{cheng94} and \cite{kowitt95}.
This flight observed an arc at a declination of 82\deg.
Preliminary results from the second flight
(MSAM1-94, 1994 June 1--2) are reported by \cite{cheng95}.
This flight duplicated the sky coverage of the first flight in order to
confirm the detection of anisotropy in the CMBR.
Analysis for the third flight (MSAM1-95, 1995 June 1--2) is in progress.
This last flight doubled our sky coverage, adding an arc at a declination of
80\fdg5 at roughly the same right ascensions.
Small improvements were made between flights, and are summarized in
Table~\ref{history}.

\begin{table}[htb]
\centering
\caption{Hardware Status Summary for the MSAM1 Flights}\label{history}
\begin{tabular}{ccl} \hline \hline
Flight & Date & Hardware Change \\
\hline
MSAM1-92 & 1992, June 4--5 & First flight of MSAM1. \\
MSAM1-94 & 1994, June 1--2 & Minimized area of the structure\\
	& & above the gondola. \\
	& & Improved coupling to balloon flight\\
	& & train (jitter mechanism).\\
	& & Tested new star camera. \\
	& & New warm preamp for the detectors. \\
MSAM1-95 & 1995, June 1--2 & New star camera. \\
	& & Improved ground screen construction \\
	& & method. \\
\hline
\end{tabular}
\end{table}

The MSAM1 configuration of the experiment was retired after MSAM1-95.
We plan to use the same telescope and gondola with a new radiometer for a
series of measurements (MSAM2) beginning in 1996.
Other instruments designed for similar observations are described in
\cite{fischer92} and \cite{meinhold93a}.

Both the gondola and the radiometer for MSAM1 have flown before.
The gondola was used with a different telescope and radiometer to map the
Galactic plane (\cite{hauser84}).
The radiometer had been flown three times for the
Far InfraRed Survey (FIRS, \cite{meyer91}, \cite{ganga93}, \cite{ganga94})
that confirmed the anisotropy detection from NASA's Cosmic Background Explorer
mission (\cite{bennett94}).
The cryostat, bolometers and filter bandpasses are discussed in
\cite{page89}, \cite{page90}, \cite{page92}, and \cite{page94}.
For MSAM1, we replaced the cryogenic chopper in the FIRS configuration
with a dichroic beam
splitter since internal referencing was not required, and we replaced the cold
feed horn and lens with an elliptical concentrator to match the telescope
optics.
The only other significant modification to the radiometer was replacing
the external preamps before MSAM1-94.
These amplifiers and associated support for the cold detector electronics
are located immediately outside the cryostat at the hermetic connector for
the signals, and provide the first gain stage for the signals.
This change improved the noise performance and stability of the detectors,
and helped to eliminate some sensitivities to microphonics and
``popcorn'' noise.

\begin{figure}[tbhp]
\begin{center}
\BoxedEPSF{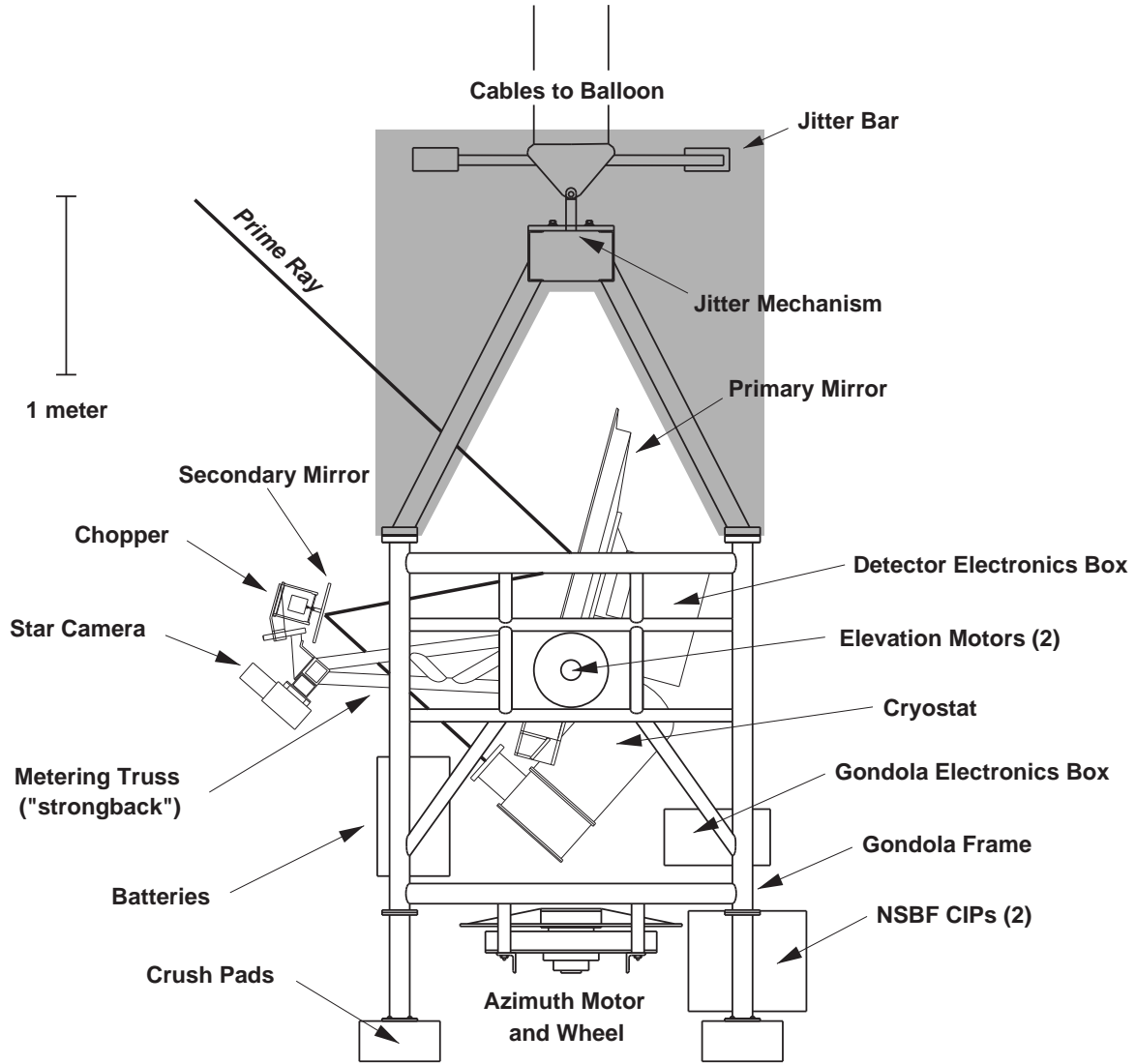 scaled 900} \\[2ex]
\end{center}
\caption{
   Side view of the gondola and telescope MSAM1-92 configuration.
   The shaded structure on top of the gondola was replaced by a wire rope
   cable system for MSAM1-94 and later flights.
   The ground shield is omitted for clarity.
}\label{gondola}
\end{figure}

The main components of the instrument are shown in Figure~\ref{gondola}.
The gondola frame
is attached to the balloon by a rigid tubular superstructure at the
top of the gondola (MSAM1-92), or a cable system (MSAM1-94 and later).
This frame supports the telemetry electronics, batteries, and the pointing
system mechanisms.

The entire gondola frame is servoed in azimuth during flight.
Two symmetrical elevation drive motors support the telescope metering
structure (the ``strongback'').
This structure holds all the elevation-controlled components.
These include: the primary mirror, the secondary mirror and chopper, the star
camera(s), the cryostat and related signal electronics, the secondary
magnetometer, and the gyroscope.

\begin{figure}[tbhp]
\begin{center}
\BoxedEPSF{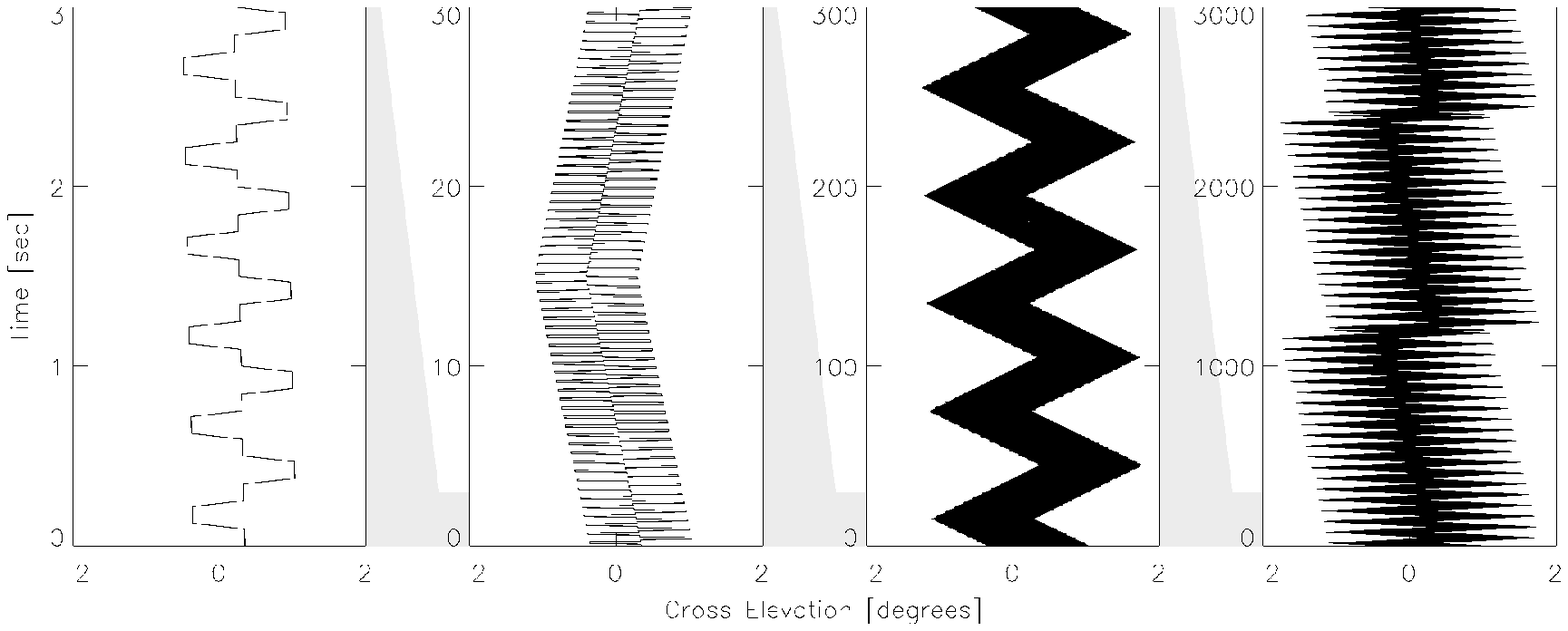 scaled 900} \\[2ex]
\end{center}
\caption{
The MSAM chopping strategy.
   Illustration of the various levels of modulation in the instrument:
   secondary mirror chopper pattern (leftmost panel),
   gondola scan pattern (middle panels),
   and sky rotation (rightmost panel).
Each panel shows the nominal position of
the telescope beam as time advances vertically;
successive panels increase the time scale ten-fold (as indicated by the
shading).
The horizontal axes plot azimuth (fixed relative to the Earth).
}\label{chop}
\end{figure}

The gondola is operated during flight in a manner that both minimizes changes
of the gondola orientation relative to the Earth, and provides three levels
of modulation to aid in the detection and removal of certain observational
errors (atmospheric drifts, telescope offset drifts, sidelobe pickup, etc.).
The first level is provided by a nutating secondary (chopper) operating at 2~Hz
(see Figure~\ref{chop}, leftmost panel).
This frequency is above the $1/f$ knee of the detectors and front-end
electronics.
The detailed operation of the chopper is described below.
The second level is the ``scan'' motion.
Here, the telescope azimuth is swept in a triangle wave pattern
with a 1 minute period and
with an amplitude matching the chopper throw.
The telescope tracks sky
inertial coordinates throughout this motion.
The third modulation level shifts the scan center approximately every
20 minutes by the chopper throw amplitude, in the direction that restores
the azimuth position to near true North (see Figure~\ref{chop}, rightmost
panel).
This has the effect of placing the scan center at the extreme position of the
previous scan, providing a large degree of sky position overlap.

In this paper, we concentrate on these items: the telescope optics,
the temperature of the optical elements,
and the attitude control systems.
Weight and power budgets are provided as these are of interest for
any balloon-borne instrument that must operate for many hours under
battery power.
This information will be useful for developing the next generation of
balloon-borne instruments for CMBR measurements
such as the
{\sl TopHat\/} telescope for the top of a balloon
(\cite{cheng93}, \cite{kowitt95a}),
or large-aperture gondolas of the standard design for high-resolution
observations.

\section{Telescope Optics}
\begin{figure}[tbhp]
\begin{center}
\BoxedEPSF{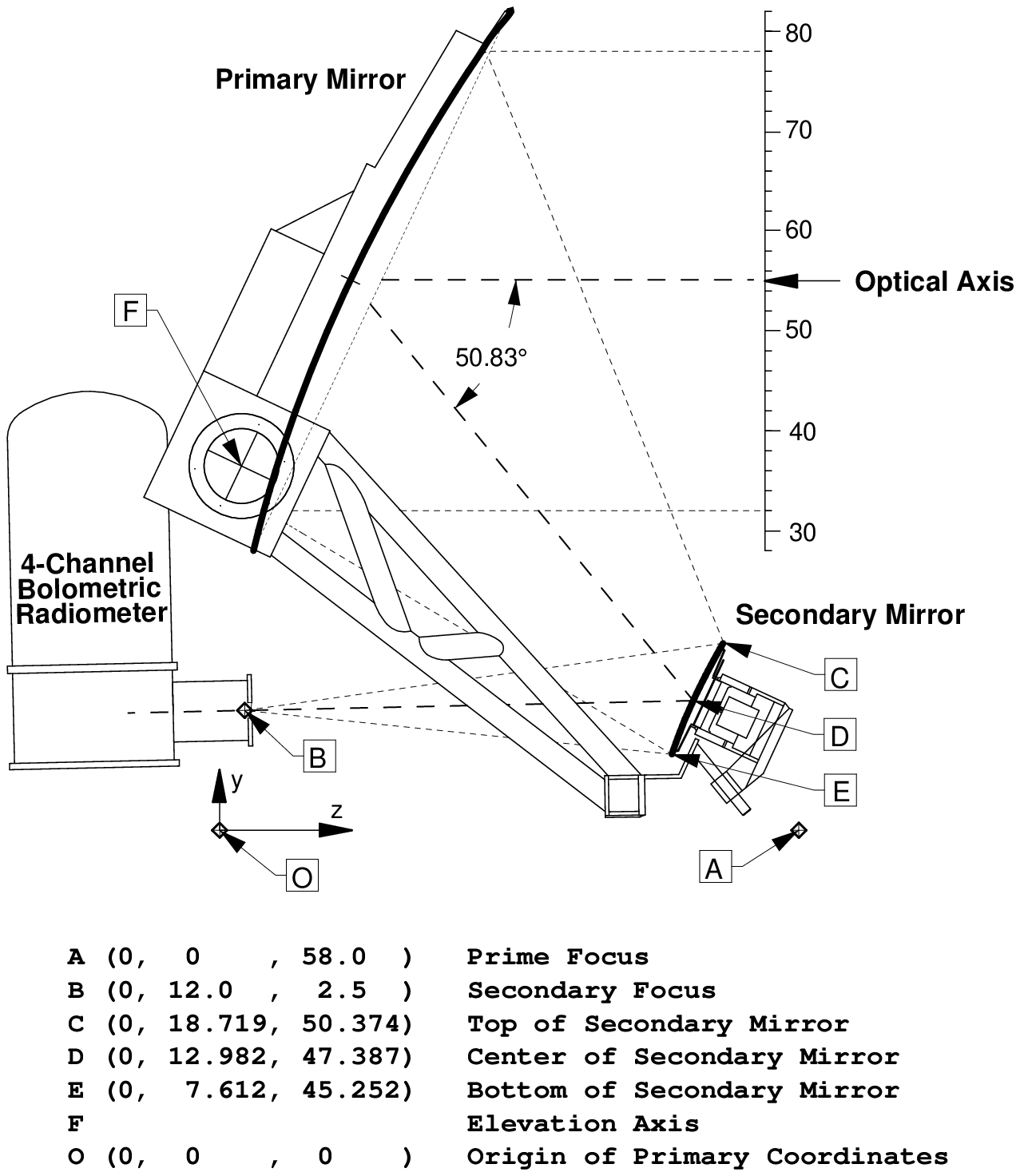 scaled 1000} \\[2ex]
\end{center}
\caption{
The MSAM telescope optics.
All distance units are in inches.
The primary mirror surface is specified by $z=(x^2+y^2)/4f$, $f=58$.
The secondary mirror surface is specified by
$z^\prime = a\protect\sqrt{1+(x^{\prime2}+y^{\prime2})/b^2}$,
$a=14.065$, $b=24.662$.
Primed coordinates have their origin centered between the prime and secondary
foci with the $+z^\prime$ axis passing through the prime focus and
$x^\prime=x$.
The chief ray is in the y-z plane.
The sky beam is parallel to the z-axis.
}\label{optics}
\end{figure}

The optical configuration is a 1372 mm, 51\deg\ off-axis Cassegrain with a
nutating secondary.
The optical layout is illustrated in Figure~\ref{optics}.
The entire telescope assembly is controlled by an altitude-azimuth mount.
The effective
beamsize\footnotemark[1]\footnotetext[1]{Effective beamsize is defined as
$2\int P d\Omega / {\sqrt{\pi \int P^2 d\Omega}}$ where $P$ is the antenna
response.}
is 37\arcmin,
and the secondary has a $\pm$ 40\arcmin\ throw.
The throw is in a plane 65\fdg24 from the beam such that 1) the center of the
chop is a line of symmetry for the optical system, and 2) the motion is
parallel to the local horizon.
While this is neither on a great circle nor at a line of constant elevation,
the small motion makes it a good approximation to either\footnotemark[2].
\footnotetext[2]{
The motion of the optical beam on the sky is approximately in the
cross-elevation direction which is defined
by a great circle perpendicular to the line of constant azimuth at the beam
center.
For an elevation $h$ and a change in azimuth
$\delta\theta$, the change in cross-elevation is
approximately $\delta\theta / \cos h$.}.
In flight, the chopper executes a four-phase cycle in the following sequence:
1) 125 ms on the the central spot,
2) 125 ms on a spot 40\arcmin\ to the left,
3) 125 ms on the central spot, and
4) 125 ms on a spot 40\arcmin\ to the right.
These times include a 23 ms transition time.
The phase of the demodulation is found by observing Jupiter,
as described in \cite{cheng94}.

This chopping cycle puts the signal information for the center
relative to the edges (double difference) at 4~Hz and its
odd harmonics, while the difference
between the two edges (single difference) is
at 2~Hz and its odd harmonics.
Thus, the double and single difference data can be analyzed separately.
These form statistically independent data sets and correspond to different
spatial frequencies on the sky.

There are several critical optical parameters that can affect the reliability
of the experiment.
First, the offset of the instrument must be small and stable to enable making
precise measurements of 30 $\mu$K temperature differences.
This is controlled mainly by the surface emissivity of the mirrors, their
thermal properties, and the symmetry of the optical paths (and thus
the chopper stability).
Additionally, the sidelobes of the telescope must be exceedingly small to
minimize sensitivity to intensity variations in the environment (outside of
the main beam).
The measured performance of the MSAM1 instrument in these areas is discussed
in subsequent sections.

\subsection{Primary and Secondary Mirrors and Chopper}
The $1372 \times 1518$~mm off-axis parabolic primary mirror has a focal
length of 1473 mm.
It is machined from a single piece of aluminum (6061-T6) and lightened by
machining 9 pockets out of the back, leaving support ribs which are 68 mm
deep and 9.5 mm wide, and a 10 mm thick ``skin" on the front surface
for a total weight of 75 kg. The lowest vibrational mode is 214 Hz.
This mirror is fabricated on a
numerically controlled mill with a ball mill.
Appropriate stress relieving after each cut preserves its dimensional
stability.
After machining, the last pass on the mill is made with a polishing tool
to remove the fine grooves left by the ball mill, leaving a surface finish
of 2 microinches.

The $276 \times 315$~mm secondary mirror is a convex hyperbola.
The secondary mirror is the limiting stop for the optical system.
The edge of the primary mirror is not illuminated, and any excess
power from the radiometer horn (at large beam angles) spills onto
the cold sky.
The secondary mirror design is similar to
the primary mirror but on a smaller scale.
Its ribs are 22 mm deep and 2.5 mm wide,
and the front ``skin" is 1 mm thick, for a total mass of 550~g,
and a moment of
inertia about the moving axis of $\sim 2.6$~g~m$^2$.
Its lowest vibrational mode is 342~Hz.

\begin{figure}[tbhp]
\begin{center}
\BoxedEPSF{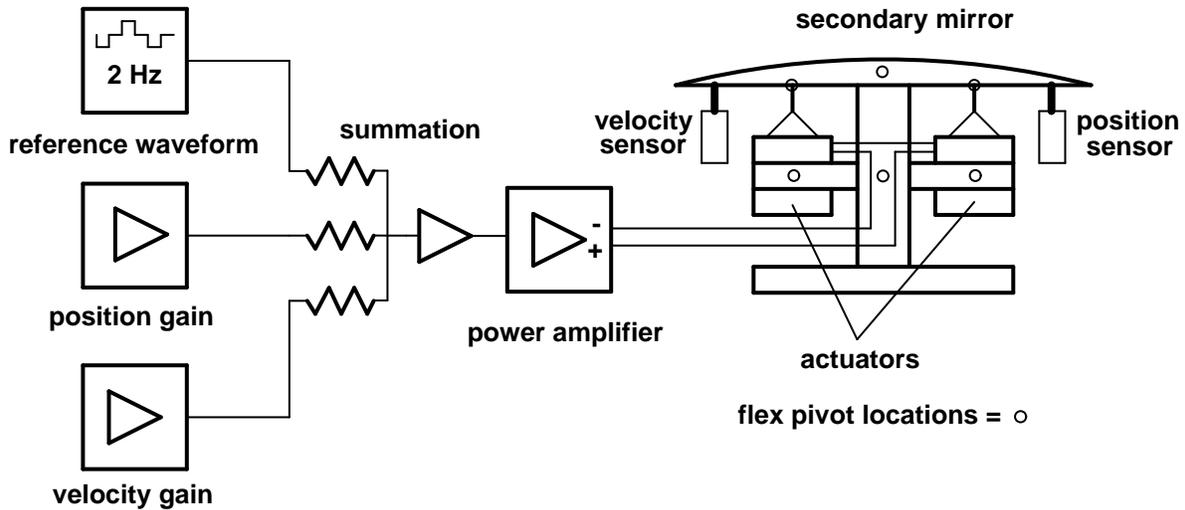 scaled 1000} \\[2ex]
\end{center}
\caption{
   Schematic diagram of the secondary mirror chopper mechanism.
}\label{chopper}
\end{figure}

To move the beam 40\arcmin\ on the sky,
the secondary mirror must be rotated  1\fdg79 about the vertical axis.
A schematic diagram of this mechanism is shown in Figure~\ref{chopper}.
It is driven by a pair of Gearing \& Watson GWV4 MkII linear
motors that weigh 2.5 kg each and together use 30~W when chopping.
Most of the power is used to deflect the centering springs in the linear
motors.
The transition time is 23~ms (to 90\%). The system is overdamped with a
$\sim 30$~ms decay time.
The drive signal is filtered to remove frequencies
outside the mirror drive response band.
It is derived from a square wave that is passed through a 4-pole
low-pass Bessel filter with a 32~Hz 3~dB point.
The chopping is stabilized by feedback from a linear variable differential
transformer (Lucas Schaevitz E 200 LVDT) operated at 16.39~kHz modulation
with a response bandwidth of 1~kHz
to position, and a linear velocity transducer
(Lucas Schaevitz 3L5 VT-Z LVT), operated with a 1.5 kHz bandwidth.
Measurements indicate the chopping is stable to 4\arcsec\
(on the sky)
over the entire flight both in the central position and in the amplitude of
the chop.
The total mass of the chopper assembly is 8 kg.

\subsection{Feed Horn}
\begin{figure}[tbhp]
\begin{center}
\BoxedEPSF{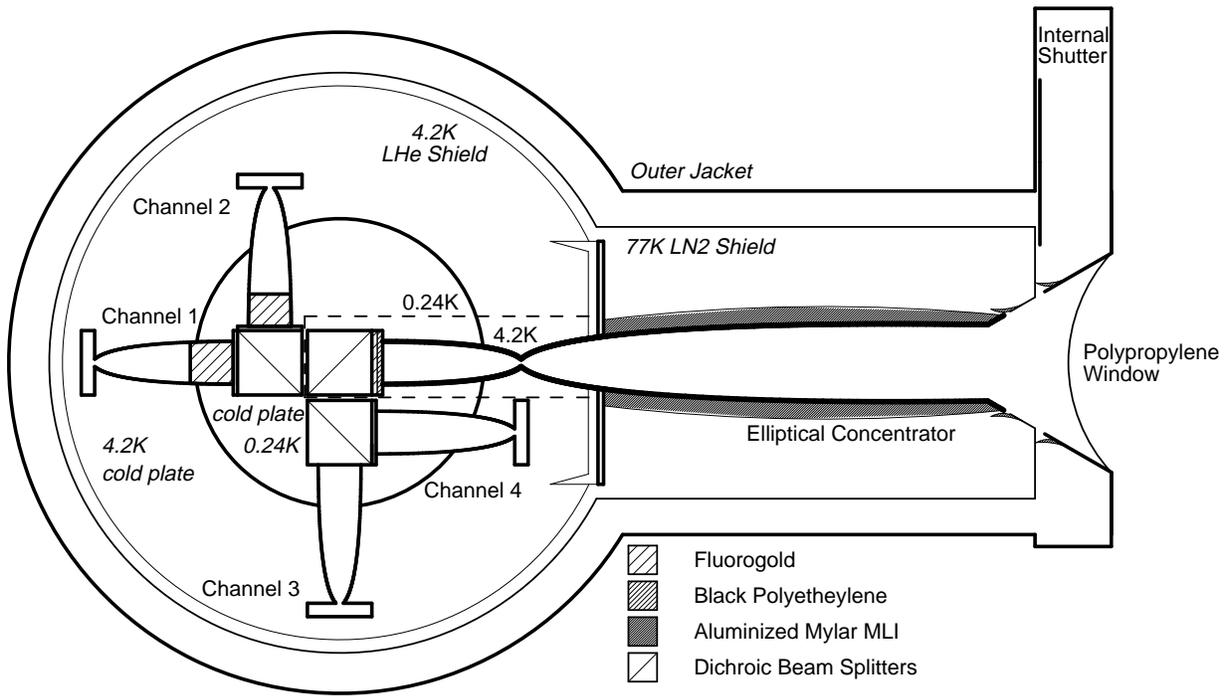 scaled 900} \\[2ex]
\end{center}
\caption{
   Schematic diagram of the cold optics in the radiometer.
   The input elliptical concentrator, its collimating horn, and the
   first optical block (with black polyethylene and fluorogold absorbers)
   are all
   maintained at close to 4.2~K.  The rest of the optics and detectors
   are at 0.24~K.
}\label{feedhorn}
\end{figure}

The beam-forming horn for the system is an elliptical Winston cone
concentrator cooled to 4.2~K (see Figure~\ref{feedhorn}).
It is designed to concentrate light from the secondary mirror
onto the feed horn throat
(4.5 mm diameter) for a design \'etendu of 0.5 cm$^2$ steradian.
The horn is 275 mm long and has a diameter
of 45.1 mm at the mouth.
A detailed geometrical ray trace shows that virtually all of the radiation
passing through the throat comes from the surface of the secondary mirror.
Diffraction effects substantially degrade this performance for the low
frequency channel.
The diffraction is estimated by
propagating 10000 rays from a point in the throat of the Winston cone to the
mouth of the Winston cone keeping track of the phase of each ray.
The pattern at the secondary mirror is then determined by integrating over
the mouth of the horn for each point at the secondary mirror.
This is repeated for each of 19 modes (at 6 cm$^{-1}$)
that the horn supports in each polarization, and the result
is the sum over all modes.
Because the E-field varies over the mouth of the horn, the final beam is
worse than simply convolving the ray trace pattern with an Airy pattern for a
45~mm circle.
The diffraction calculations agree reasonably well with ground tests and
laboratory measurements.
Both show that $\sim 10$\% of the radiation is from outside of the
secondary mirror
for the 5.6 cm$^{-1}$ channel.
This fraction is less than 4\% for the shorter wavelength channels.

\subsection{Radiometer}
The radiometer design and its support electronics are essentially the same
as that used for the FIRS (\cite{page89}, \cite{page94}).
The main difference is the replacement of the cryogenic chopper with a
dichroic at the first optical block (see Figure~\ref{feedhorn}) and
the new feed horn discussed in the previous section.

The optical structures are made from copper (OFHC 101).
These blocks support all the filter elements, provide thermal sinking to
the appropriate cold stage, and enclose the
25.4~mm (1~inch)
diameter light pipes.

The incoming beam is collimated and blocked with black polyethylene at
4.2~K, and then split twice using capacitive grid dichroics (once at 4.2~K
and once at 0.24~K) to form the four channels.
Each channel has a Winston concentrator feeding a monolithic silicon bolometer
(\cite{downey84}).
In front of each concentrator is a band defining filter.
The detailed passband measurements are discussed in \cite{page94}.

The bolometer outputs are fed into cold JFET followers, and then to
preamps external to the cryostat.
The internal wiring of the high impedance leads is through small-gauge
manganin wire that is glued inside a small, thin, stainless steel tube in
order to minimize microphonic pickup and to provide shielding.
This arrangement creates a low thermal conductivity, rigid, coaxial signal
line.

All of the internal cryostat lines, including housekeeping, power, and buffered
detector signals pass through connectors manufactured by Microtech, Inc.,
which are potted in RF absorbing epoxy (Emerson-Cumming CR-117).
These connectors are compact and reliable at cryogenic temperatures.
However, their small size makes assembly somewhat challenging, especially
when used with small-gauge stainless steel or manganin wires.

All the cryostat lines are also fed through RF $\pi$-filters at the hermetic
connectors on the cryostat shell.
This precaution has allowed us to make repeatable noise measurements in
different environments (including at the National Scientific Balloon
Facility, NSBF, where there are frequently
intentional RF transmissions of modest power).

\subsection{Alignment}
With the two mirrors and the feed horn,
there are a total of 12 degrees of freedom
involved in the alignment process.
Each optical element is placed at its nominal design position by a machined
mount, and then shimmed by measurement.
In combination these mounts allow four degrees of freedom:
the horn can be rotated about the radiometer's vertical axis, it can
be raised or lowered, the secondary mirror can be moved closer or further from
the primary and it can be rotated about its vertical axis.
The telescope is aligned by first centering the feed horn beam on the
secondary mirror by mapping the feed horn pattern at the secondary mirror
location and shimming the position of the cryostat.
Subsequently, the position of the secondary mirror is shimmed so that the
center position beam is symmetrically placed on the primary mirror.

\begin{figure}[tbhp]
\begin{center}
\BoxedEPSF{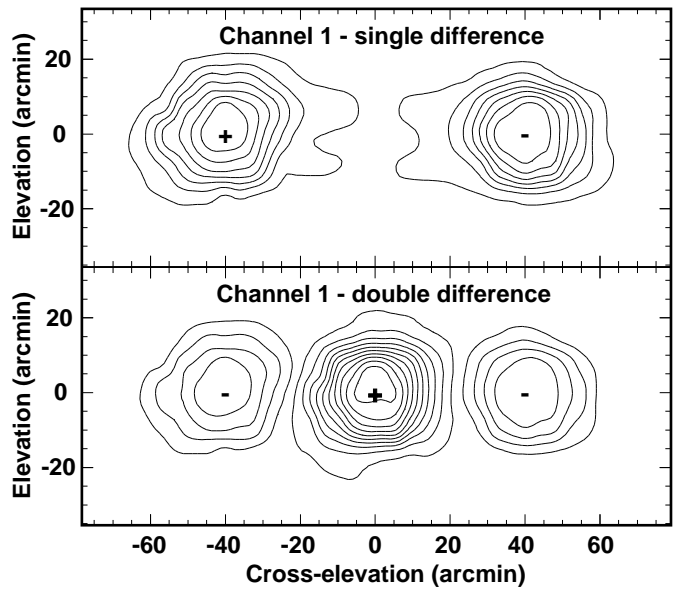 scaled 500} \\[2ex]
\end{center}
\caption{
   Single and double difference maps made in flight on Jupiter.
   The contours are at uniform linear intervals.
   The other channels are very similar.
These data were taken by moving the gondola in a raster pattern, 1\fdg1
peak-to-peak in elevation and 3\deg\ in cross-elevation.
The elevation motion is in 9 steps, while the cross-elevation motion is a
smooth scan.  The raster represents approximately 10 minutes of data.
}\label{beammap}
\end{figure}

The mainlobe of the beam was measured by observing both a blackbody source
and a microwave source on the ground.
The measured beam profile is used to verify the alignment of the
telescope and to establish a reference position for the
IR beam center in the CCD star camera.
The final determination of the beam pattern and IR beam center position
is made in-flight by observing Jupiter.
The results of one such determination are shown in Figure~\ref{beammap}.
The ground-based maps agree with the flight measurements in the central
lobe to 10\% of the peak response, limited by the accuracy of the ground
test data.
There is typically a shift of approximately 10\arcmin\ in
cross-elevation and 5\arcmin\ in elevation in beam position
(relative to the camera) between the ground and flight measurements.

The large beam size makes alignment of the telescope relatively
straightforward.
Typical alignment tolerances for the various adjustments are
$\sim 2$~mm for the cryostat position and $\sim 10$~mm for the centering
of the beam on the primary mirror.

\subsection{Ground Shield}\label{groundshield}
\begin{figure}[tbhp]
\begin{center}
\BoxedEPSF{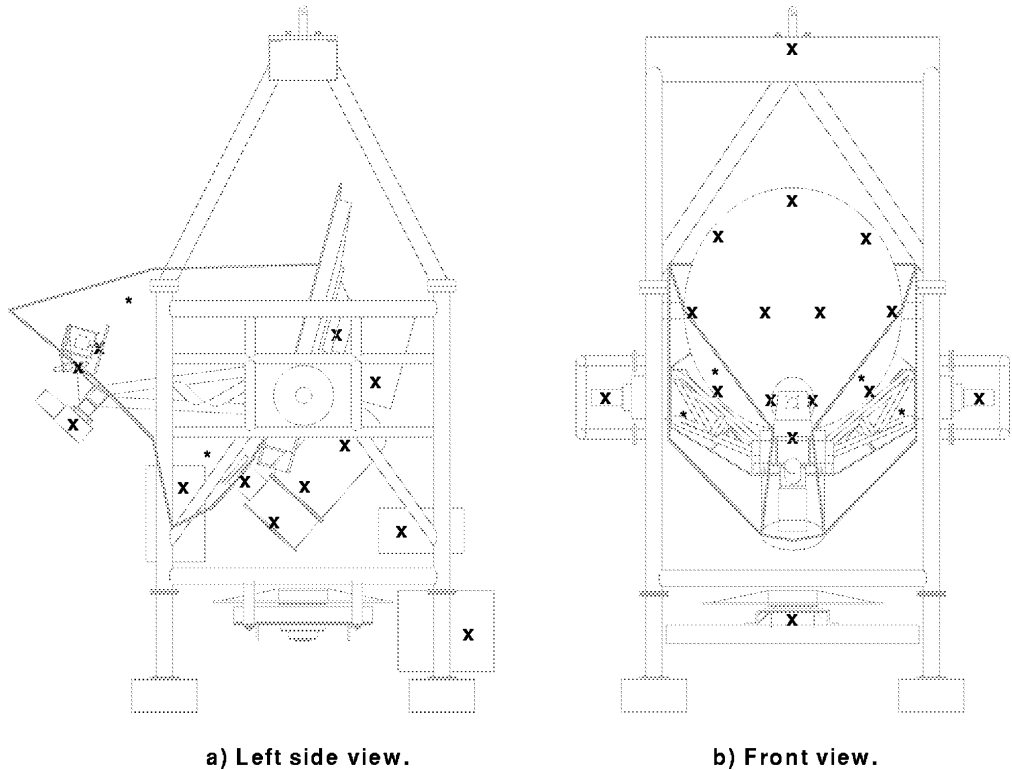 scaled 800} \\[2ex]
\end{center}
\caption{
   Two views of gondola showing some thermometer locations. Many items
   have been omitted for clarity.
   The ground shield outline is shown.
   The *'s show the approximate positions of the temperature sensors on the
   ground screens, and the X's show the sensor locations on the other
   gondola components.
   Note that the shield hides the horn, most of the primary mirror, and
   the secondary mirror from the horizon and below.
   Diagram a) shows the gondola near the nominal observing position.
   In b) the elevation of the telescope is at a lower position to
   afford a clearer view of the primary mirror.
}\label{thermometers}
\end{figure}

The ground shield is a critical element that shields the
radiometer input aperture,
the secondary mirror,
and most of the primary mirror from the Earth during observations.
The shield is constructed at the launch site from foam
building insulation.
For MSAM1-92, a 15 m$^2$, 50 kg shield was
constructed with foiled, polyisocyanurate fiberboard building insulation
covered on the  interior side with an additional 0.25~mm aluminum sheet
(see Figure~\ref{thermometers}).
With the wire rope top structure for MSAM1-94, a smaller 5~m$^2$ shield
was built using similar methods.
Unfortunately, some of this foam exploded during ascent and at
float altitude.
The additional aluminum sheet remained, however, preserving the optical
integrity of the shield.
For MSAM1-95, we used unbacked closed-cell styrofoam with an aluminum sheet
on the interior side, and aluminized mylar on the outside.
These materials withstood the rigors of the (night) flight environment
without incident.
We performed some destructive daytime tests at the end of the
flight to gain some experience for future packages.
Styrofoam covered with aluminized mylar melts with only a few tens of
minutes of exposure to sunlight if the aluminized side faces out.

Two sets of tests were performed to measure the far sidelobe response of
the telescope.  The first, a near-field test, provides a sensitive
measurement of the beam response to sources near the horizon, and
helps to pinpoint ``trouble-spots'' in the shielding and gondola
structure that could potentially cause unacceptable levels of response.
The second, a far-field test, maps the response of the beam
as a function of angle from the mainlobe as the telescope
is raised in elevation.

The near-field test was done with the gondola
placed in a parking lot, pointed at its nominal observing elevation
of $\sim 42^\circ$.  A mm-wave source, placed on a tripod $\sim 5$~m
from the gondola, was pointed successively at the body of the gondola,
the top edge of the telescope shielding, and the gondola top support
structure.
This procedure was repeated at 10 positions on
a circle surrounding the gondola
(at 0, 10, 45, 90, 135, 180, 225, 270, 315, and 350 degrees azimuth).

For the MSAM1-92 flight, the near-field test was performed using an
Alpha Industries 180V 60.6 GHz diode
followed by a Millitech MU3W15-01 tripler
(10 $\mu$W) and a standard 21 dB horn used as the point source.
Although this 182 GHz (6.07 cm$^{-1}$) source could only be seen in
the lowest frequency channel, this channel has the largest fraction of stray
radiation and the most diffraction, suggesting that it would have the
largest effect.  We found that the
solid top structure reflected some radiation
($\sim -55$~dB) into the detector,
motivating the change to the cable top structure
for the later flights.

For the second and third (MSAM1-94, MSAM1-95) flights we used a 150~GHz Litton
LST9423XX Gunn diode (30 mW) and a 21 dB horn
as the point source for the near-field tests.
With the new top structure, which was smaller and farther
from the beam axis than the old structure,
we found no positions with sensitivity above $-80$~dB.

The far-field sidelobe test was done just prior to the second flight
(MSAM1-94).  A 1-meter on-axis Cassegrain telescope ($\sim 53$~dB gain,
the ``broadcast" telescope)
was placed on top of the 25~m control tower at the NSBF
in Palestine, Texas, with the MSAM gondola
situated 740~m away in a clearing (the ``old launch pad'').
The same 150~GHz Gunn diode
used for the near-field tests was coupled to the broadcast telescope
through a Winston concentrator at the secondary focus.

\begin{figure}[tbhp]
\begin{center}
\BoxedEPSF{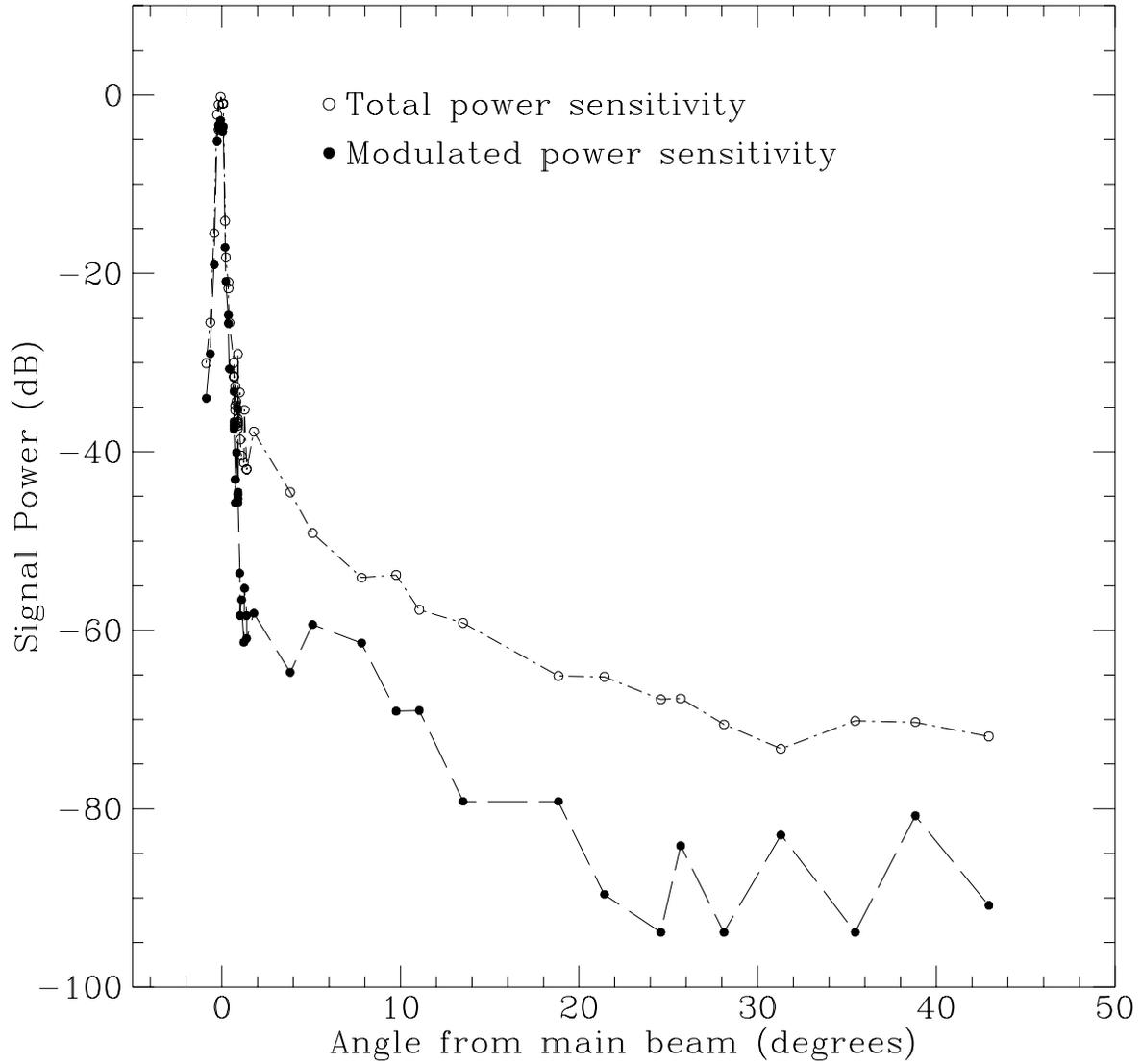 scaled 800} \\[2ex]
\end{center}
\caption{
   Far sidelobes as measured before the MSAM1-94 flight using a 1-meter
   telescope for illumination.  Both the total power sensitivity and
   the sensitivity to the component that is synchronous with the
   secondary mirror modulation are shown.  The latter is typically a factor
   of 10 below the total power sensitivity.
   The plotted data are for a scan in the elevation direction.
   The horizontal axis shows the angle of the transmitter {\sl below}
   the main beam in elevation.
}\label{sidelobes}
\end{figure}

The broadcast telescope was pointed at the MSAM telescope and
left in that position throughout the tests.  To avoid saturating
the bolometric detector in MSAM while measuring the
on-axis response, it was necessary
to attenuate the broadcast power
by inserting microwave absorbing material in the feed optics.
After measuring the on-axis response (with the
MSAM telescope pointed at a low elevation near the horizon)
the MSAM telescope was raised in elevation to map the response
as this moved the source off axis.  Once the MSAM telescope was
pointed a few degrees off axis
the absorbing material was removed from the feed optics.
The removal of the material
caused a change in MSAM's response, which in turn calibrated
the attenuation factor of the material.
This procedure provided sufficient dynamic range
to map the sidelobes to roughly -100~dB,
as shown in Figure~\ref{sidelobes}.  The uncertainty in
the level of the far sidelobes relative to the main beam
is roughly 3~dB, dominated by the uncertainty in calibrating
the attenuating material.

The same broadcast telescope fed by a
$\sim 1000$~K blackbody source allowed us to map
the mainlobe in all four channels out to
$\sim -25$~dB.  This was used as a check for beam symmetry
and focus prior to flight.  A similar arrangement,
but using a small (0.1~m) polyethylene lens instead of the
1~m telescope, was used for the same purpose prior to the
first and third flights.  This setup was also effective
for checking the mainlobe out to $\sim -20$~dB.

Another test before the MSAM1-94 flight measured
the telescope response at elevation angles higher than the beam.
The MSAM telescope was placed $\sim 12$~m from the edge of the
control tower, and illuminated by the 150~GHz Gunn diode and
21~dB horn from the top of that tower.
This measurement constrains the instrument response
at the edge of the balloon during nominal flight observations
to be less than -55~dB.
The measurement is dominated by reflections off
the side of the control tower, and is thus an upper limit.

For all of these measurements,
the source was modulated at $\sim 0.7$~Hz,
while the secondary was moving in its normal 3-beam pattern
(at 2~Hz full cycle).  We recorded both the fundamental modulation
frequency as well as the
appropriate sidebands of the secondary mirror chopper frequency,
thus measuring both the chopped (synchronous
with the secondary mirror modulation) and total power sensitivities.
All the numbers quoted in the text are for the chopped case unless otherwise
noted.
The total power response is roughly a factor of 10 higher than the
chopped response.

\subsection{Sidelobe Performance}
\begin{figure}[tbhp]
\begin{center}
\BoxedEPSF{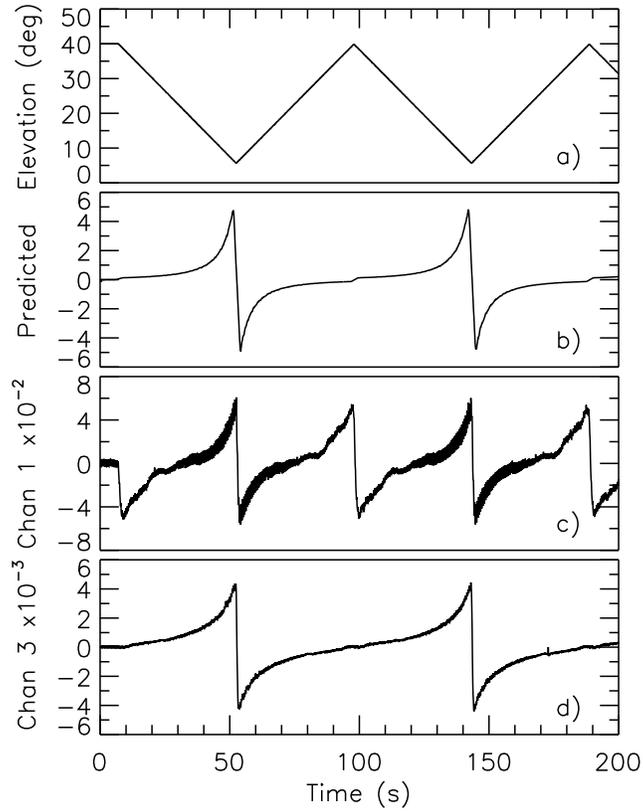 scaled 500} \\[2ex]
\end{center}
\caption{
   Data from the elevation scans for MSAM1-92
   showing a diffraction effect in channel 1.
   This effect was eliminated for subsequent flights.
   All plots are a function of time measured in seconds: a) elevation
   in degrees,
   b) predicted signal shape, c) signal in channel 1,
   d) signal in channel 3.
   The feature at $t=172$ in (d) is a cosmic ray hit on the detector.
The bottom three panels are plotted in linear power units
on the vertical axis.
The signals are AC coupled, causing the diffraction effect to appear
as a double-cusped signal in channel 1.
}\label{elscan}
\end{figure}

In-flight elevation scans are made to check obvious sidelobe sensitivities
as well as the total atmospheric emission.
The results of these scans for MSAM1-92 are shown in Figure~\ref{elscan}
(after replacement of the top structure of the gondola, the subsequent
flights do not show any such effect).
The expected emission from the atmosphere is $T_z \csc \theta$,
where $T_z$ is the zenith temperature and $\theta$ is the zenith angle.
Since the bolometers are AC coupled, we expect to see the derivative
of $T_z$ as in the channel 3 data.
An anomalous feature appears in channel 1 at high elevations, probably
arising from stray radiation from the top of the gondola.
The antenna temperature from this source is $\sim 4$~K,
an amount that is consistent with $\sim 4$\%\footnotemark[3].
\footnotetext[3]{
We measured the sidelobe response while illuminating the gondola from 30
positions on the ``horizon''
during ground testing.
These positions included all of the intuitively obvious worse-case
locations.
Two of these had a -55~dB response.
}
of the solid angle subtended by the ground having
the $-55$~dB response seen in the test.
For MSAM1-94 and later flights, the large feature at high elevation is no
longer present,
consistent with the hypothesis that it was a result of reflection off the
top of the gondola.

There are three different ways that such an effect might contaminate
the data.
One is the signal induced by elevation motions ($\sim 3$~mK/arcmin).
While the elevation should be roughly independent of time for our
observations,
there may be some residual coupling between the elevation and azimuth
positions arising from small misalignments of the gyroscope axes.
Ground-based measurements place an upper limit of 6 arcmin on the misalignment
of the axes, which corresponds to an upper limit of 0.18 arcmin tilt
over the gondola scan distance (1\fdg5 peak-to-peak).
The net result is a total change in observed sky power of $\sim$~540~$\mu$K.
The secondary mirror chopping reduces this effect by at least a factor
of 10 (confirmed by the elevation scan data), and the
strategy of repeated measurements of the same point in the
sky in the right, center and left chopped beams reduces the effect further
by at least a factor of 2.
The net result from this source is conservatively estimated
to be less than 50 $\mu$K.

A second contamination source is present even if the elevation stays constant.
The ultimate source of the radiation (presumably the ground) is not constant,
varying both in time (as the balloon floats along) and in space. The
spatial correlations are not well understood as we have only a rough idea of
the far sidelobe pattern, but tens of $\mu$K signals are possible.

A third contamination source could arise from small optical misalignments
that show up as an inconsistency in results when the gondola is moving
towards the left or the right.
We have tried to detect these effects, and the difference is less
than 15 $\mu$K (95\% confidence) for all channels combined.
This result shows that
any such bias is much smaller than the primary sky signal
and can be safely ignored.

In light of some of these uncertainties, the MSAM1-94 flight covered the
same sky as the MSAM1-92 flight in order to validate the observations.
Both flights concentrated on an arc at $\delta=81$\fdg8 and
$15<\alpha<20$~hours.
Preliminary analysis suggests the two flights detect very similar structure,
placing the best limit on any contamination to either flight
(\cite{kowitt95}).

\subsection{Shutters}
A commandable shutter is included inside the cryostat vacuum jacket
immediately before the feed horn.
This is a copper sheet with an evaporated gold surface that covers the
entire open aperture of the horn.
When closed, the shutter presents a $\sim 160$~K surface to the feed horn.
This shutter is opened before launch and kept open for most of the flight.
Near or at sunrise, it is closed to block the optical signal, but the
gondola continues to execute normal scanning motion for 20 minutes.
This test provides a useful confirmation that observed offsets are optical.

A backup external shutter mechanism is available for protecting the
polypropylene window at the cryostat entrance aperture.
This shutter places
an aluminum sheet between the cryostat window and the incoming beam,
and prevents a severe problem should the telescope inadvertently point
at the Sun after the attitude control system is powered down.
Since we need this protection only after the flight is over, the
mechanism is a one-shot, spring-deployed design that is actuated by
melting a pair of nylon lines.

\section{Attitude Control}
The telescope consists of the primary mirror, the secondary mirror,
a 100 kg cryostat, and 45 kg bolometer electronics all mounted on a 140 kg
supporting structure (strongback).  The two axis active control system,
with damping, has feedback from a number of sensors to control pointing.
The system can be organized by the control axis and sensor
(see Table \ref{table:pointing}).
The roll axis (and thus rotation about the optical axis) is passively
controlled by the balance of the gondola.
Although the balance adjustment is simple (lead bricks),
we are able to level the gondola to 10\arcmin\ before launch.
In flight it is stable to 3\arcmin\ after the swinging attenuates,
($\sim 30$~minutes after float altitude is reached).

\begin{table}
\centering
\caption{Pointing System Sensors}\label{table:pointing}
\begin{tabular}{rccc} \hline
Axis: & Roll & Pitch & Yaw  \\
\hline\hline
 & (left-right tilt) & (elevation) & (azimuth)  \\
first sensor: & secondary gyro* & primary gyro & primary gyro  \\
second sensor: & magnetometer & angle encoder & magnetometer  \\
third sensor: & camera & inclinometer & magnetometer  \\
fourth sensor: & --- & camera & camera  \\[2pt]
first control: & none & elevation drive & reaction wheel  \\
second control: & --- & fluid pump & jitter bearing  \\
\hline
\end{tabular}
\\
\flushleft
*Only on MSAM1-92 and MSAM1-95.\\
\end{table}

\begin{figure}[tbhp]
\begin{center}
\BoxedEPSF{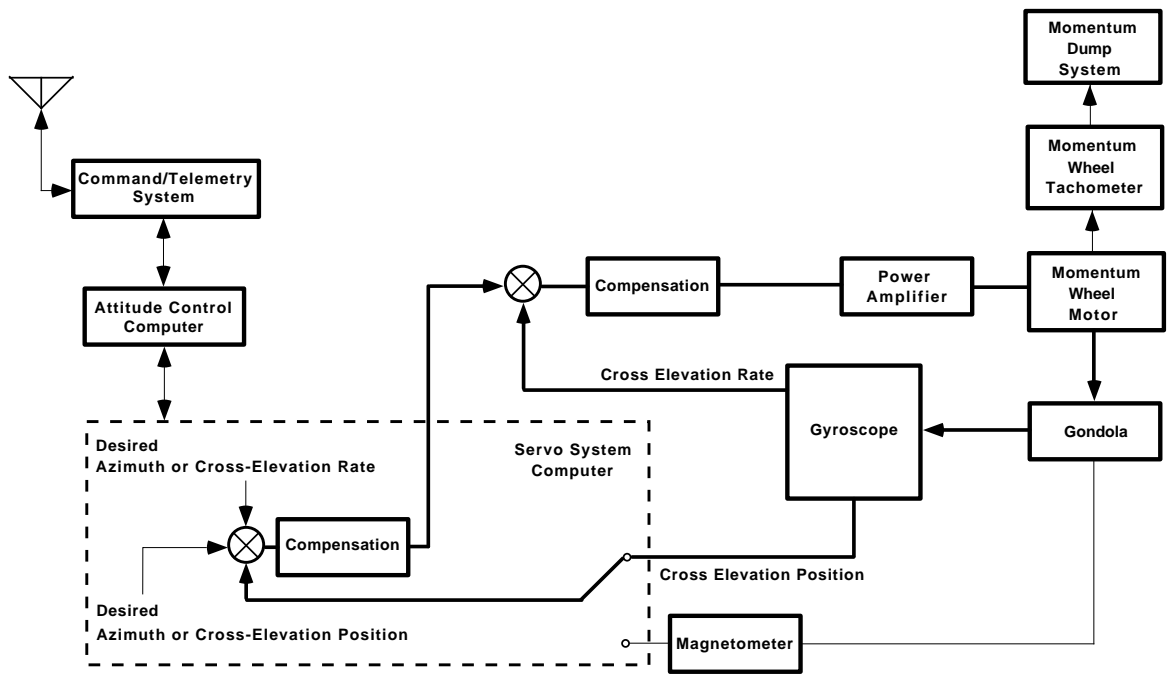 scaled 1000} \\[2ex]
\end{center}
\caption{
   Block diagram of the azimuth attitude control system.
   The switch illustrates
   a configuration change that is available during a flight.
   The indicated switch setting is the nominal flight configuration.
   The elevation control system is somewhat simpler because a momentum
   dump is not required, but is of a similar design.
}\label{servosystem}
\end{figure}

There are a number of sensors and controls for each of the axes.
Much of this redundancy is for historical reasons and would not be necessary
in a new design.
However, all of the sensors have proven to be valuable for detailed
reconstruction of the in-flight telescope performance.
A block diagram of the control system is shown in Figure~\ref{servosystem}.

\subsection{Pendulation}
\begin{figure}[tbhp]
\begin{center}
\BoxedEPSF{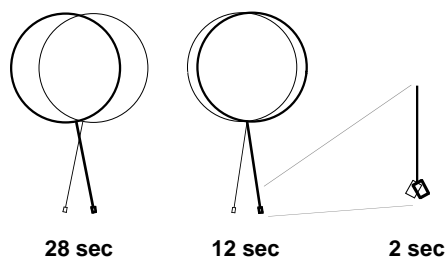 scaled 600} \\[2ex]
\end{center}
\caption{
   Illustration of the three pendular modes of the balloon/gondola system.
   Displacements are exaggerated.
}\label{pendulation}
\end{figure}

The major mechanical flight components are: a 1880 kg gondola,
a 2300 kg balloon,
a 200 kg ladder line,
and 720 kg of helium gas in the balloon.
Assumptions used in the calculations include:
a rigid spherical balloon with uniform mass
distribution over the surface (130~m diameter), a rigid 104~m ladder line
with uniform mass distribution,
and a rigid gondola with a radius of gyration of 1~m.
Prediction of the various modes of pendulation is relatively simple,
the only
difficulty is calculating the amount of air that is effectively coupled to
the  balloon,
which was varied from 0 to 5100 kg (the mass of air displaced by the
balloon). There are three important modes (see Figure~\ref{pendulation}).

The lowest frequency mode is one in which the entire flight assembly swings
back and forth as a unit.  The predicted period for this mode is between
26 and 34 seconds depending on the coupling to the air.  In the early part of
the flight an oscillation was observed with a period of 28 seconds which
damps in about 30 minutes.
Using the model described above, this corresponds to a coupled air mass
of 15\% of the total displacement of the balloon.

The next lowest mode is one in which the gondola and flight train swings in
one direction as the balloon rotates in the opposite direction.
The calculated period for this mode is between 11 and 14 seconds
depending on the balloon to air coupling.
In the early part of the flight an oscillation was observed with a period of
12 seconds, consistent with a 5\% coupled air mass.

For the next mode, the gondola swings as the balloon and flight
train rotate in the opposite direction.  The calculated period for this
mode is 2 seconds. The main uncertainty is the mass distribution of the
gondola.  There is an observed oscillation with a period of 2 seconds
in the early part of the flight and after the gondola makes large motions
(for example, when we made a  90\deg\ azimuth rotation to observe the Coma
cluster).

We conclude that the three detected frequencies are manifestations of the
three predicted modes and that other possible modes are either not excited
or quickly damped.
These effects are visible only after wind-driven excitation during ascent
and after a major slew of the telescope (e.g., to locate a calibration
source).
Their amplitudes immediately after such excitation is $\sim 10$\arcmin,
but they damp out on timescales of 10 minutes and
do not affect telescope pointing during the data taking portions of
the flight.

\subsection{Gyroscopes}
The main gyroscope is a Teledyne Systems Model SDG-4\#8005304-503 two axis
rate gyroscope mounted on the strongback with the gyroscope axis aligned with
the
telescope optical axis to within 1\deg.
The gyroscope is accurate to $1'$ with a drift rate less than
$10'$/hour after warmup and compensation. The electronics use analog
components to integrate the signal and provide the primary
pointing reference over 20-minute scan periods.

For MSAM1-92, the secondary gyroscopes are a pair of Kearfott vertical
gyroscopes  (model C704132001) which derive their long term stability from
gravity.
These are used outside of the control loops to measure the swinging of the
gondola.
No secondary gyroscopes were flown on MSAM1-94.
On MSAM1-95, we used a surplus Pershing II, two-axis
rate gyroscope (model 11502525-039)
as the secondary gyroscope to monitor roll,
with the goal of using it in the control loop for future flights.

\subsection{Magnetometers}
Two flux gate magnetometers are flown on the package to measure the local
magnetic field.
The primary magnetometer is a two-axis device that is fixed to the gondola
frame.  The secondary magnetometer senses three orthogonal axes and is
mounted on the primary mirror support structure.
Both devices are a flux gate design.  The primary magnetometer was built
by Goddard, the secondary magnetometer was purchased from Schoensted.
The accuracy limitations in this system are the distortions of the
Earth's magnetic field by the electronics and the magnetic materials in the
gondola itself.
The absolute accuracy of the magnetometers is limited to about 1\deg\ but
their low weight and power consumption ($<$1 kg, $<$1 W),
combined with their absolute azimuth readout, make them attractive sensors.

\subsection {Camera}
The ultimate pointing reference for MSAM1-92 is a charge injection device
(CID) star camera
(General Electric model 2200, 6 kg, 20 W).  The $128\times128$ detector
has a field of view of $8\deg\times8\deg$ at a resolution of
$3\farcm8\times3\farcm8$.  The camera is sensitive to
wavelengths from 400 to 950~nm. With an integration time of 10 seconds, it
detects F stars of $M_V<$~5.6 and K stars of $M_V<$~5.8.  On-board
processing of the images identifies the location and brightness of the
stars in each frame.  Because of the long integration time, the telescope
must be held fixed to take a picture. This is done between scans
(every 20 minutes).
Only the star positions are transmitted to the ground.

For MSAM1-94, we added a second star camera (Photometrics model
Star~1, 3 kg, 15~W).  This $576\times384$ charge coupled device (CCD)
detector has a field of view
of $9\deg\times6\deg$ at a resolution of $\sim 0\farcm9$. This camera is
sensitive to wavelengths from 400 to 1000 nm.
With an integration time of 0.1 seconds,
it easily  detects stars of $M_V\sim 7$.
Although the required integration time is short
enough to take pictures at any time,
the temporal registration between the camera and gyroscope data
was not good
enough to use these pictures except when we stopped to get a picture with
the CID camera between each 20 minute scan.

For MSAM1-95, we used only the Photometrics camera, and refined the time
registration scheme so that readings taken during scans can be used for
pointing reconstruction.

\subsection{Elevation Control}
The elevation, or pitch control is maintained by a pair of torque motors
(Inland Motors model T-7203, 33 kg, 40 W peak power each).
An error signal is obtained by subtracting the desired angle from the sensor
angle (the sensor can be selected from the list in
Table~\ref{table:pointing}).
This error signal is fed back with an adjustable gain and offset to the
torque motors to correct the error.
When the gain is too high a mode involving the pitching of the
telescope and counter-pitching of the gondola is excited.
With proper adjustment, the servo has a time constant of 0.3 seconds and is
stable to 2\arcmin, more than adequate for the 37\arcmin\ beam.

The long term stability of the pitch is perturbed by the evaporation of
liquid nitrogen from the cryostat.
This is compensated by a system that pumps fluid between
two tanks, one on the cryostat support structure, and the other on the
gondola frame.
The pump speed is determined by the averaged elevation motor current, thereby
working to reduce any elevation torque imbalance.

\subsection{Azimuth Control}
The azimuth, or yaw, control is similar to the elevation control.
The azimuth is maintained by a torque motor identical to the pitch motors
driving a momentum wheel (77 kg, 15~kg-m$^2$) to change the angular momentum
of the gondola frame.
Mechanical and/or electrical biases would tend to spin up the momentum wheel
so the angular momentum is transferred to the flight train and eventually to
the balloon through the ``jitter'' mechanism described below.
The system is similar to the one discussed in \cite{hazen74}
and \cite{hazen74a}.

\begin{figure}[tbhp]
\begin{center}
\BoxedEPSF{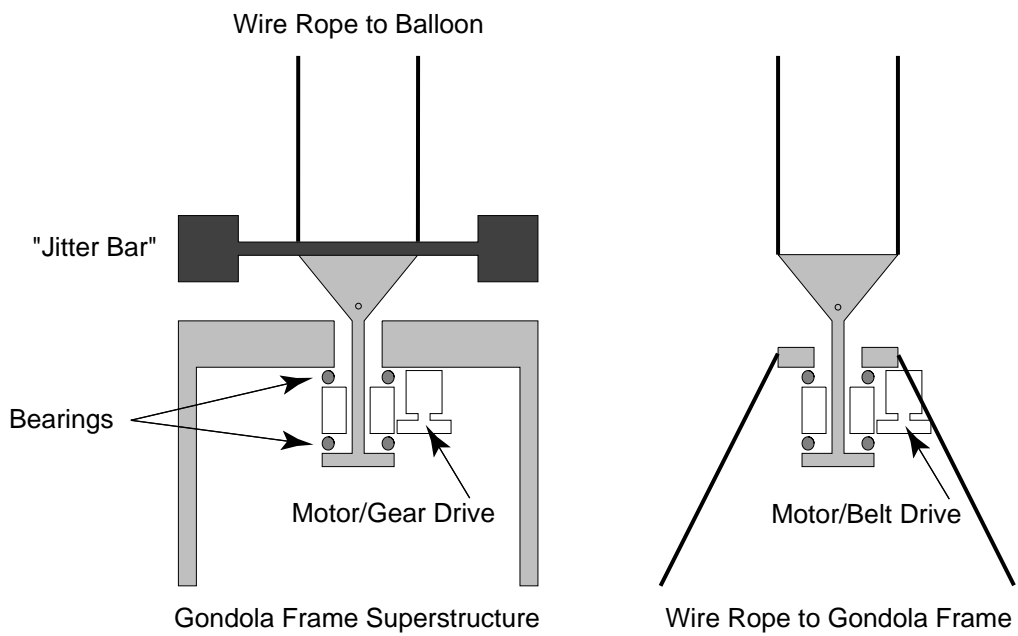 scaled 1000} \\[2ex]
\end{center}
\caption{
   Left: schematic diagram of the jitter mechanism used for MSAM1-92.
   Right:
   schematic diagram of the corresponding improved mechanism for MSAM1-94
   and later flights.
}\label{jitter}
\end{figure}

For MSAM1-92, the entire gondola was hung from a pair of bearings
suspended by the flight train (see Figure~\ref{jitter}a).
The jitter mechanism keeps the bearings in motion by oscillating
a short shaft between the two bearings, thus maintaining dynamic friction
rather than the higher static friction (stiction).  The stiction is
a factor of 5 higher than the dynamic friction and the transition introduces
an undesirable nonlinear factor in the control system.
A jitter bar (34 kg, 11 kg-m$^2$) is added to the bottom of the flight train
to provide reaction inertia for
the bearing connected to the flight train rather than induce
a twist in the flight train (43 Nt-m/rad).

The jitter mechanism also transfers angular momentum to the
flight train.  This is done by biasing the motion of the bearing to the
direction necessary to apply a torque (0 to 0.5~Nt-m) to the flight train.
The jitter mechanism (including motor) weighs 55~kg and uses 28~W.

For MSAM1-94 a new jitter mechanism (Figure~\ref{jitter}b) was installed.
The bearings for this new jitter mechanism had a factor of $\sim 10$
less friction,
and we were able to run without the oscillation and without the jitter bar.
The bearings used are from Precision Consolidated Bearings, model 51212P/5.

The adjustment of the azimuth feedback is complicated by two factors.
The first is the lack of information about the flight train dynamics.
The bottom of the flight train is a pair of parallel cables, 0.48~m
apart, and the upper part is connected to a parachute and then to the balloon.
Neither can be treated as a rigid structure.
The length of the flight train (104 m) complicates efforts to develop a
ground-based test setup.
In addition, the different air-damping effects on the ground and at float
altitude need to be taken into account.

The second factor is that although the secondary magnetometers measure
true azimuth, the gyroscope measures cross-elevation change which is
$\sim \delta\theta / \cos h$, where $\delta\theta$ is the azimuth change
and $h$ is the elevation.
Hence the gain (and damping) in the feedback loop is a function
of elevation when using the gyroscope.
This could be corrected by the flight software but was not,
since we generally operate over a small range of elevations.
The azimuth time constant is 2 seconds and a 0\fdg7 step function
settles to 90\% settling in 1.5 seconds.
In flight, the azimuth error is 2\arcmin\ rms with long term drifts of
$\sim 4$\arcmin.

\subsection{Camera Results}
\begin{figure}[tbhp]
\begin{center}
\BoxedEPSF{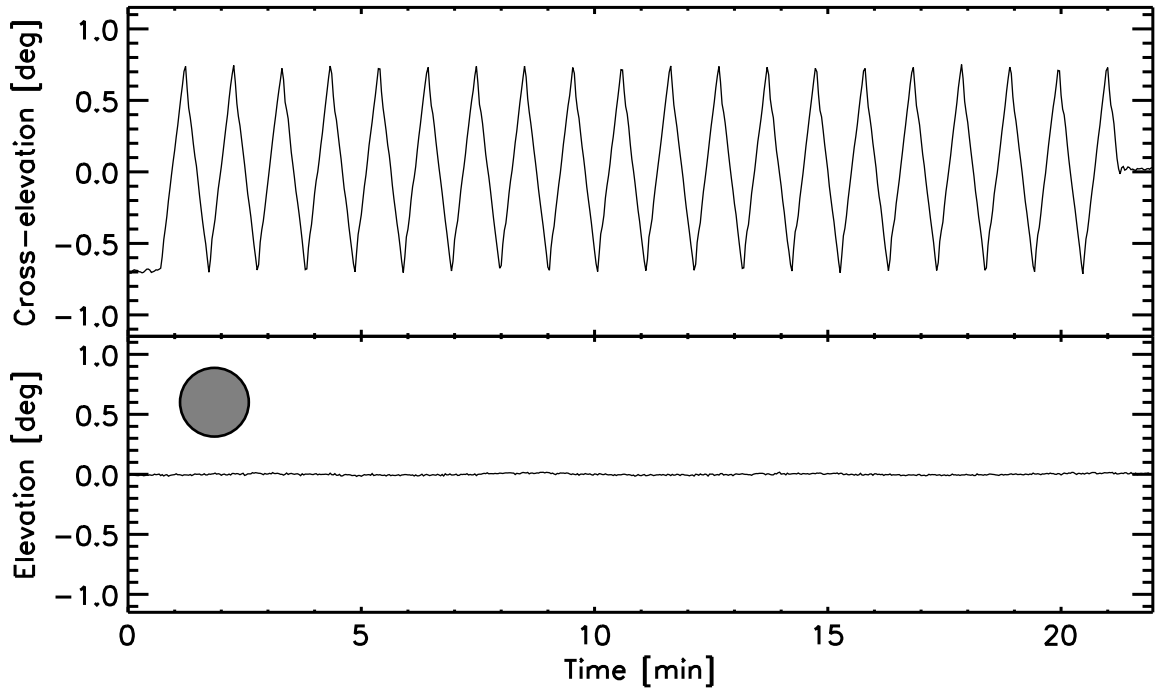 scaled 1000} \\[2ex]
\end{center}
\caption{
   In-flight elevation and cross-elevation during a 20-minute scan
   (as measured by the gyroscope integrated position signal).
   These coordinates are fixed relative to the sky.
   This pattern is repeated by joining the flat regions of cross-elevation
   at the edges of the plot.
   Both plots are on the same scale, the zero-points are arbitrary.
   The circle shows our nominal beam size.
}\label{elandaz}
\end{figure}

Absolute pointing is determined by matching images taken by the star
camera against a star catalog.
This fixes the position of the camera frame at the time the exposure
was taken.
Between pictures, position is interpolated using the gyroscope,
with a small linear drift correction to
make the gyroscope readings consistent with the camera pictures.
This correction is typically 2\arcmin\ in 20~minutes.
The relative position of the camera frame and the telescope beam is
fixed by a simultaneous observation of Jupiter with the camera and the
telescope/radiometer.
The resulting pointing is accurate to 2\farcm5, limited by
the gyroscope drift correction.
A plot of the MSAM1 pointing performance during a typical 20-minute scan
is shown in Figure~\ref{elandaz}.

\section{Thermal Considerations}
The temperature of the ambient optical surfaces is a critical issue since
they contribute a significant amount of the total power seen by the detectors
($\sim$~50\% in the lowest
frequency channel, and increasing to $\agt~90$\% for the highest frequency
channel).
Numerous temperature sensors are included to monitor critical components
for correlating with the final data.
Most of the optical surfaces are close to the ambient temperature because
the MSAM flights are single night flights.
Significant thermal perturbations are encountered as the Sun rises.

For ambient components, Analog Devices AD590 temperature sensors
are used to measure their temperature to 0.1 K with an absolute accuracy
of $\sim 1$~K.
These thermometers are glued onto the back surface of the mirrors and other
locations on the shield and gondola (see Figure~\ref{thermometers}).
There are 10 thermometers on the primary mirror, 2 on the secondary mirror,
4 on the ground shield, and 12 in other locations on the
gondola frame.
One thermometer (elevation motor temperature) failed at launch for MSAM1-92
and one thermometer on the secondary mirror failed on ascent for MSAM1-94,
probably due to ground shield deformations interfering with the wiring.

Float altitude for the balloon is $\sim 38$~km, where the
air temperature is $\sim 235$~K.
The minimum temperature during ascent occurs at the tropopause
($\sim 18$~km) where it is $215 \pm 2$~K.
Since the pressure is low (2.25 torr or 300 Pa),
equilibration to the ambient
temperature happens slowly, especially for massive objects like the
primary mirror and the gondola frame.

\begin{figure}[tbhp]
\begin{center}
\BoxedEPSF{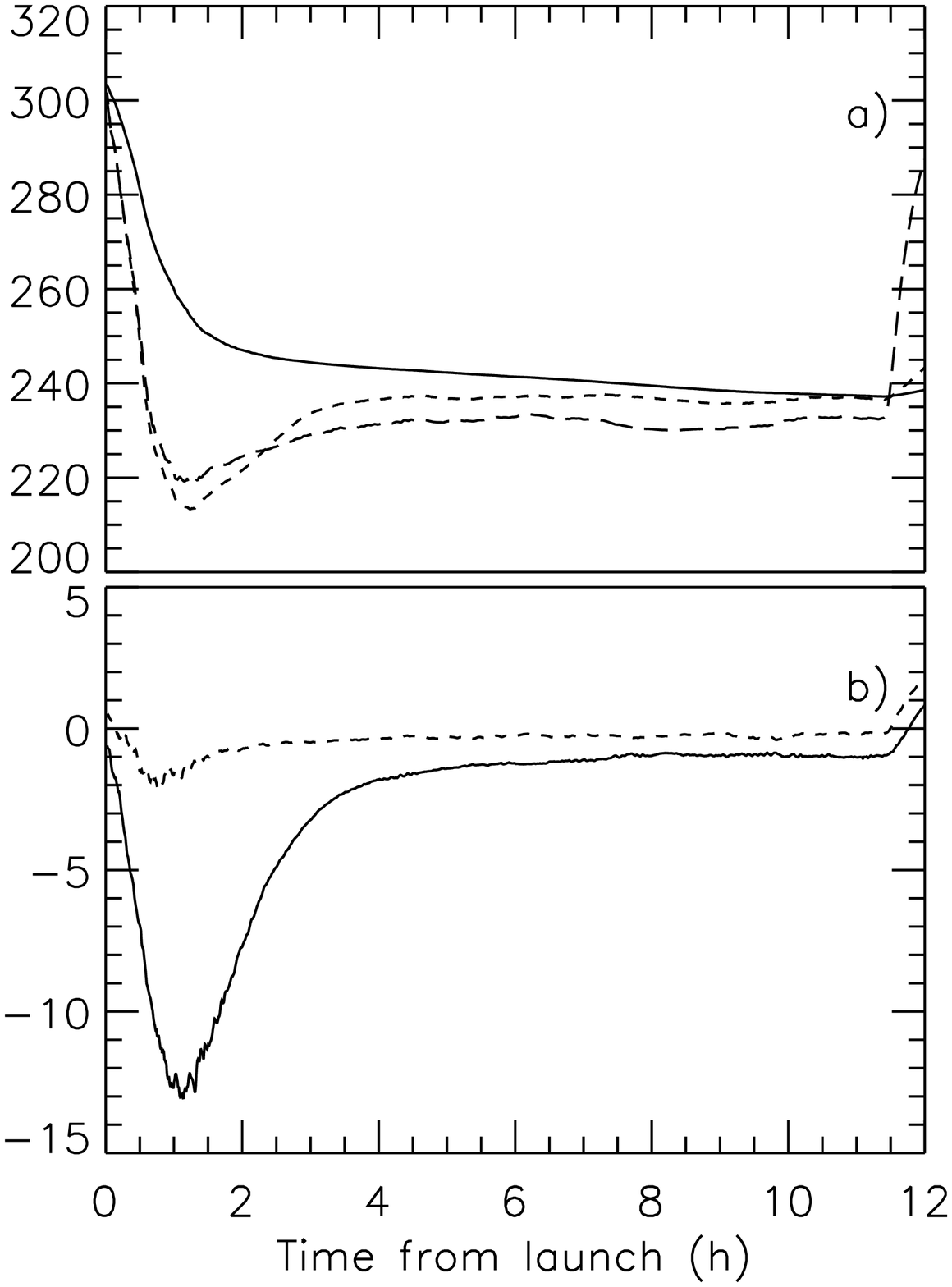 scaled 500} \\[2ex]
\end{center}
\caption{
   a) Temperature of the primary mirror (solid line),
   secondary mirror (dashed line), and shield (long dashed line).
   The kinks at 11.4 hours are caused by sunrise.
   b) The temperature difference between the primary mirror bottom and top
   (solid line)
   shows the cooling effect of the ascent and the warming effect of the
   Sun on the top.
   The side to side difference in primary mirror temperature
   (dashed line) shows much smaller effects.
   The zeros in both a) and b) have not been calibrated and may be in
   error by $\sim 0.5$~K.
   The absolute temperature data in a) have a resolution of 0.1~K.
   The temperature difference data in b) have a resolution of 1~mK.
   During a typical flight, CMBR data is not taken until 3 hours after
   launch.
}\label{primarytemp}
\end{figure}

The primary mirror starts the flight at the ground temperature and is cooled
as the gondola ascends.
Cooling is assisted by the $6 \pm 2$~m/s downward wind during the ascent.
The cooling is faster on the more exposed top half of the mirror
than the bottom half.
At float altitude this cooling slows and it does not
come to full equilibrium even after 6 hours (see Figure~\ref{primarytemp}).
During the observation period,
the cooling rate is roughly uniform at 1.25~K/hr.
The temperature gradients diminish much faster, the longest time constant
being $\sim 30$ minutes for the smooth gradient over the whole mirror.
More complex patterns have shorter time constants.

The temperature of the primary mirror is critical since even the
0.5\% emissivity of
good aluminum surfaces radiates more power than the CMBR in our lower
frequency bands and much more in the two higher bands.
For anisotropy measurements, only the variation (both in time and in position)
is potentially harmful to the experiment,
but the required sensitivity is 5 orders of magnitude below the total power.
There are two different effects to consider,
one involving a uniform temperature and a varying
emissivity over the mirror surface, and
the other involving a nonuniform temperature.

Assuming a uniform temperature over the mirror but variations of $\sim 0.05$\%
in the emissivity gives signals of order 120 mK.  These signals are averaged
over large sections of the mirror and $\sim 90$\% of the area observed is
the same in all three beam positions.  This will reduce the estimated signal
to $\sim 1$ mK, which is the expected offset.
In fact, in MSAM1-94, we see a very stable offset that is between 1 and 6~mK,
depending on channel.
The excess seems to be due to one
of the positions of the secondary mirrors reflecting part of the gondola
into the detectors.
The temperature of the mirror only changes by 8~K over
the observing time which will produce a drift of 30~$\mu$K.
Offset drifts slower than roughly 2.5 minutes
are removed by a spline fit
in the analysis, reducing this effect in the final data to below
our sensitivity limit.

Assuming a uniform emissivity of 0.5\%, but a 1~K temperature gradient
across the mirror (the largest seen during the CMBR observations)
gives a potential signal of 4~mK.
This signal diminishes with a characteristic time of 30 minutes.
The observation strategy repeats the scans every minute, reducing the signal
to 100 $\mu$K.
Averaging scans reduces this further and the result is again smaller than the
observation noise.
The exact amount of rejection depends on ability of the drift
removal algorithm to separate the slow drift of the mirror cooling
from the 1 minute scans of the telescope.
Higher modes in the thermal structure of the primary mirror are less well
attenuated but the effects are smaller and the coupling to the beam is less
efficient.  Thus, these modes are estimated to have a smaller net effect.

Although the secondary mirror radiates as much power onto the detectors as
the primary mirror, the coupling of this power into the modulated signal
is not significant.
The modulation of the emitted power is only $\sim 0.05$\% (from the
geometric change in the size of the secondary mirror as seen by the
detectors, and from the small modulation in emissivity due to a change
in the angle of incidence).
For a $\sim 0.5$\% emissivity of the mirror surface, the mirror will
generate a $\sim 1$~mK offset.
Thus, even a 10~K change in the secondary mirror
temperature will generate only a 25~$\mu$K change in the modulated signal
before additional attenuation by the other two levels of modulation.
The temperature of the secondary mirror during a flight is also
more uniform than the primary mirror,
as shown in Figure~{\ref{primarytemp}}.

\section{Telemetry and Commanding}
Monitoring and controlling the observations are done through a remote
commanding and telemetry system.  The
NSBF supplies a package, called
the Consolidated Instrument Package (CIP),
which is flown on the payload, along with associated ground
equipment which provides the experimenter with the following command
and telemetry capabilities:
\begin{enumerate}
\item Several telemetry channels of various bandwidths.  The widest channels
have maximum bandwidths of 49.5~kHz, 10.5~kHz, and 4.5~kHz.
A signal which meets the maximum bandwidth and amplitude requirements is
fed to the CIP, and the NSBF's ground equipment reproduces this signal
on the ground.
\item 80 discrete commands, also called relay commands.
The experimenter directs that one of these
commands be sent by sending a packet to the NSBF's ground computer.
Sending a relay command causes a particular digital line from the CIP
to be pulsed.  Relay commands can be sent at a rate of about 2 per
second.
\item A digital word channel.
These also are sent by sending a packet to the ground computer.
They appear at a strobed parallel interface on the CIP.
16 bit digital words can be sent on
this channel at a rate of about 1 per second.
\end{enumerate}

\begin{figure}[tbhp]
\begin{center}
\BoxedEPSF{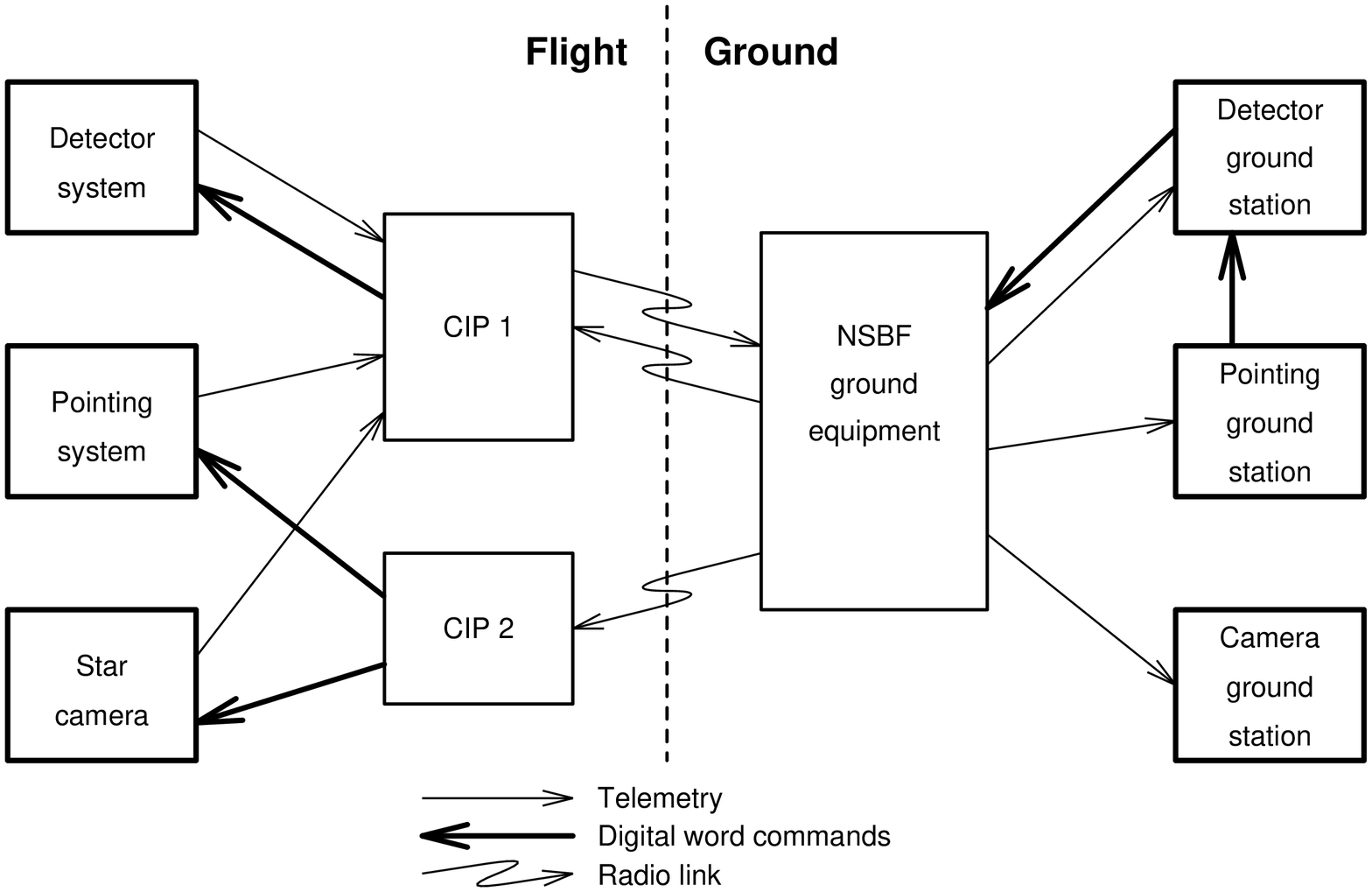 scaled 600} \\[2ex]
\end{center}
\caption{
Block diagram of the command and telemetry system.
The discrete commands are not shown; on the ground, they follow the same
paths as the digital word commands, and on the gondola, each subsystem
uses some of the discrete commands from the corresponding CIP.
The blocks with the lighter boundaries represent NSBF supplied equipment.
}\label{tm}
\end{figure}

The telemetry and command layout of this instrument reflects its
origin as two separate experiments.  To integrate the systems as
easily as possibly, we have largely preserved their original telemetry
and commanding interfaces.  As a result, we use the commanding
systems of two CIPs (though the telemetry transmitter of only one).
Figure~\ref{tm}
shows a diagram of the MSAM1 telemetry and commanding systems.

In the various configurations for the three flights this instrument
uses 3 or 4 telemetry channels.  The system that controls the pointing
of the gondola produces a biphase stream that includes information
about the pointing sensors and actuators and the state of the
controlling programs.  This stream has a bit rate of 20.48~kHz.
The system that controls the detectors produces a biphase stream at
4.096~kHz, including the detector data and information about the
cryostat, chopper, etc.  Finally, the camera systems produce telemetry
containing information about star locations.  The old camera system,
which was flown on the first two flights, produced intermittent
biphase packets at 800~Hz.  The new camera system, which was
flown on the second and third flights, produced a RS-232 stream at
4800 baud.

The digital word channel of one CIP communicates with the detector
system.  Each transmission sets one of 16 registers in the detector system;
the high-order four bits select which register, and the low-order twelve bits
specify the value to be set.

The digital word channel of the other CIP communicates with the
pointing system and the camera systems.  These systems use a packet
of words to specify each command.  The packet encodes which system the
command is directed to, the action to be performed, any needed
parameters for the action, and framing and length information.

The relay commands of the CIPs are used primarily for turning on and
off latching relays.  Relays are used to control batteries and power
supplies, and to turn on and off various parts of the instrument,
e.g., the liquid helium level sensor in the cryostat and the
telescope balance pumps.

Neither the commanding nor the telemetry channels of the CIP are
reliable in the communication theoretical sense; in other words, the
system as supplied does not detect transmission failures and
automatically resend.  Typical failures are:
\begin{enumerate}
\item Dropouts in the telemetry channels, usually for a fraction of a
second.
\item Failure of relay commands to arrive.
\item Digital word failing to arrive, or arriving altered.
\end{enumerate}
In some cases, transmission can be quite poor, particularly for
sending commands.  Our experience with this has impressed upon us the
necessity of keeping the length of command
packets as short as possible, and adding automatic detection and retry
of transmission failures.

The unreliable nature of the relay commands is easily dealt with by
a general design rule: make the effect of sending a relay command
twice the same as sending it once.  For example, if a relay command is
used to turn on a particular relay, then sending the command twice
also has the effect of turning on the relay.  Then, for critical commands,
one adopts a practice of sending the command several times.
Note that ``toggling'' type commands are strictly avoided.

Noise in the telemetry downlink can be partially compensated for by an
error detection and correction scheme.
For the radiometer data stream, we have used a hardware
ECC generator (designed for computer storage disk subsystems)
to encode the redundant information.
Software in the ground system performs error correction where possible and,
at a minimum, flags unreliable data.  For a normal flight with ``good''
telemetry, this system typically corrects several hundreds of dropouts
per flight, and flags on the order of one hundred hard dropouts.

\section{Power Systems}
The power systems for the radiometer and the gondola are also independent.
Radiometer electronics and heater power is supplied
by two 32~V lithium SDX battery packs
(model B7901-12 from Eternacell, Inc.) supplied by the NSBF.
These are augmented by six 15~V packs (model B9525) for several small
motors and relays.
A PowerCube, Inc. modular DC-DC converter (part number XDD1784)
is used to generate the
supply voltages, and has a primary conversion frequency of $\sim$~30~kHz.
Because of the potential sensitivity of the detectors to high frequency
power,
the supply voltages for the detector preamps and the bias lines for the
detectors have an additional level of regulation from a set of linear
regulators, and are then heavily filtered.
We have not been able to measure any undesirable effects from using
the DC-DC power converters in this way.
During ground testing where the batteries are replaced by high current
DC-DC power supplies, a small effect at the conversion frequency of
18~kHz can be measured on the preamp power lines.
This noise is attributed to some unavoidable ground currents, and is
not present in the flight configuration.

Gondola power is provided by 15 sets of 32~V model B7901-12 packs.
Eight of these are used for the motors.  We have sized the
battery capacity for the maximum current draw of the motors, resulting
in a capacity that is four times the average over an entire flight.
The gondola electronics uses 7 packs, and powers the gondola command,
control, telemetry and star camera functions.

Secondary gondola voltages are
generated by a set of Vicor DC-DC converters whose chopping frequencies are
not detectable in the radiometer output signals.
All power supply clocks are operated at random phase with respect to the
chopped optical signal.

\section{Weight and Power Budgets}
Table~\ref{weightandpower} summarizes the overall weight and power
budgets for the instrument.
\begin{table}
\centering
	\caption{Summary of Weights and Power}\label{weightandpower}
	\begin{tabular}{rcccc} \hline \hline
	   Item & MSAM1-92  & MSAM1-94 & MSAM1-95 & Power\\
            & Weight    & Weight   & Weight   & Consumption\\
		    & (kg) & (kg) & (kg) & (W)  \\
		Primary Mirror: & 75 & 75 & 75 & - \\
		Secondary Mirror: & 8 & 8 & 8 & 30 \\
		Ground Shield: & 50 & 15 & 15 & - \\
		Cryostat/Detectors: & 105 & 105 & 105 & 0.1 \\
		Strongback: & 140 & 140 & 140 & - \\
		Primary Gryo: & 15 & 15 & 15 & 36 \\
		Secondary Gryo: & 10 & - & 10 & 48 \\
		Elevation Mechanism: & 66 & 66 & 66 & 80 \\
		Azimuth Mechanism: & 110 & 110 & 110 & 40 \\
		Fluid Mechanism: & 20 & 20 & 20 & 10 \\
		Jitter Mechanism: & 55 & 20 & 20 & 28 \\
		Radiometer Electronics: & 45 & 45 & 45 & 90 \\
		Gondola Electronics: & 30 & 40 & 40 & 120 \\
		Camera \& Electronics: & 16 & 26 & 10 & 20 \\
		Power Conditioning: & 15 & 15 & 15 & 80 \\
		Batteries: & 110 & 120 & 130 & * \\
		Frame: & 450 & 380 & 380 & - \\
		Telemetry (NSBF): & 50 & 85 & 85 &** \\
		Lead Bricks: & 50 & 60 & 30 & - \\
		Ballast: & 320 & 320 & 320 & - \\
		Crush Pads: & 40 & 40 & 40 & - \\
		Misc: & 100 & 105 & 91 & - \\ \hline
		TOTAL: & 1880 & 1810 & 1770 & 585 \\ \hline
	\end{tabular}
\\
\flushleft
* The batteries supply 20 KW-Hr. \\
*** Power for telemetry transmission supplied by an internal lithium battery
in the CIP.\\
	\protect\label{table:summary}
\end{table}
The reduction in weight of the gondola frame and jitter mechanism
after MSAM1-92 is a result of switching to the cable top structure.
The gondola electronics weight changes as we upgrade various subsystems.
The star camera weight increased for MSAM1-94 because we flew both the
old CID and new CCD cameras.
We increased the battery complement for each flight in order to
accommodate tests for new subsystems.

\section{Conclusions}
The pointing performance of the MSAM telescope has been superb over the
MSAM1 series of flights.
The accuracy and stability of the platform is largely responsible for
simplifying operational procedures, contributing significantly to the
success of the project.
The reliability of this 15 year old gondola was initially a concern, but
by upgrading a small number of selected subsystems,
it has operated almost flawlessly for the MSAM1 flights.
At most, 1 to 2 hours of flight time was lost (out of almost 40 hours) due
to gondola related anomalies, including difficulties with
the command link that accounted for most of this loss.

The optical performance after the improvement in the gondola superstructure
is verified to be adequate for the measurements.
However, the sensitivity of CMBR measurements to Earthshine continues to
be an area of concern because of the difficulty of providing adequate
shielding and the concomitant validation measurements.
The method we have described for mapping the sidelobe performance of
the gondola is strongly recommended for all similar payloads
to help alleviate some of these concerns.

\section{Acknowledgments}
We would like to thank the staff of the National Scientific Balloon
Facility in Palestine, Texas for continuing to make ballooning a
satisfying and productive experimental activity.
In particular, for this project, they took on an increased burden by
agreeing to provide two CIPs for our package so that we can carry out
the measurements
with a minimum of electronic changes to the gondola and radiometer.
The National Aeronautics and Space Administration
supports this research through grants
NAGW 1841,
RTOP 188-44,
NGT 50908,
and
NGT 50720.

\end{document}